\documentclass{article}
\usepackage{graphicx}
\usepackage{subfigure}
\usepackage{wrapfig}

\usepackage{url}  
\usepackage[utf8]{inputenc}  
\usepackage{subfigure}  
\usepackage{latexsym}  
\usepackage{amsmath} 
\usepackage{amssymb} 
\usepackage{algorithm}
\usepackage{algorithmic}
\newtheorem{definition}{Definition}  
\newtheorem{proposition}{Proposition}

\newtheorem{property}{Property}

\newtheorem{theorem}{Theorem}  

\newenvironment{proof}{\noindent {\it Proof.}}{$\Box$\vskip1ex}

\graphicspath{{figures/}}

\begin{document}

\title{\vspace*{-2cm}Faster and Simpler Minimal Conflicting\\ Set
  Identification\thanks{This work is partly supported by the french MAPPI project
    (ANR-2010-COSI-004).}}

\author{Aida Ouangraoua\thanks{INRIA, Centre de recherche INRIA
    Haute-Borne, Bât. A, Park Plaza 40 avenue Halley, 59650 Villeneuve
    d'Ascq, France.  {\tt aida.ouangraoua@inria.fr}} \and Mathieu
  Raffinot\thanks{CNRS/LIAFA, Universit\'e Paris Diderot - Paris 7,
    France, {\tt raffinot@liafa.jussieu.fr}} 
}

\maketitle

\begin{abstract}
Let ${\cal C}$ be a finite set of $n$ elements and ${\cal
  R}=\{r_1,r_2, \ldots , r_m\}$ a family of $m$ subsets of ${\cal
  C}$. A subset ${\cal X}$ of ${\cal R}$ verifies the \emph{Consecutive Ones
Property (C1P)} if there exists a permutation $P$ of ${\cal C}$ such
that each $r_i$ in ${\cal X}$ is an interval of $P$. A \emph{Minimal
Conflicting Set (MCS)} ${\cal S} \subseteq {\cal R}$ is a subset of
 ${\cal R}$ that does not verify the C1P, but such that any of its proper 
subsets does. In this paper, we present a new simpler and faster algorithm 
to decide if a given element $r \in {\cal R}$ belongs to at least one MCS. 
Our algorithm runs in $O(n^2m^2 + nm^7)$, largely improving
the current $O(m^6n^5 (m+n)^2 \log(m+n))$ fastest algorithm of
[Blin {\em et al}, CSR 2011]. The new algorithm is based on an alternative
approach considering minimal forbidden induced subgraphs of interval
graphs instead of Tucker matrices.
\end{abstract}

\section{Introduction}

Let ${\cal C} = \{c_1, \ldots ,c_n\}$ be a finite set of $n$ elements
and ${\cal R}=\{r_1,r_2, \ldots , r_m\}$ a family of $m$ subsets of
${\cal C}$. Those sets can be seen as a $m \times n$ 0-1 matrix
$M=({\cal R}, {\cal C})$, such that the set ${\cal C}$ represents the 
columns of the matrix, and the set ${\cal R}$ the rows of the matrix:
each $r_i\in {\cal R}$ represents the set of columns where row $i$ has
an entry 1.




A subset ${\cal X}$ of ${\cal R}$ verifies the consecutive ones
property (C1P) if there exists a permutation $P$ of ${\cal C}$ such
that each $r_i$ in ${\cal X}$ is an interval of $P$. Testing the
consecutive ones property is the core of many algorithms that have
applications in a wide range of domains, from VLSI circuit conception
through planar embeddings \cite{nr-pgd-04} to computational biology
for the reconstruction of ancestral genomes
\cite{BBCC2004,Stephane2010,CHSY2009,Chauve08,SW2009}. 
We focus on this last field in this paper.


On real biological matrices, the C1P is rarely verified, and only some
subsets of rows might verify the desired property.
However, the combinatorics of such sets is difficult to
handle, and a strategy to deal with them has been proposed
in \cite{BBCC2004,Chauve08,SW2009}. It consists in identifying
the rows belonging to minimal conflicting subsets of rows that 
do not verify the C1P, but such that any of their
row subset does. 

\begin{definition}
A set ${\cal S} \subseteq {\cal R}, {\cal S} \neq \emptyset$ is a {\em Minimal
  Conflicting Set} (MCS) if ${\cal S}$ does not verify the C1P, but such 
that $\forall {\cal X},{\cal X} \subset S$, the set ${\cal X}$ verifies the C1P.
\end{definition}



\begin{wrapfigure}[15]{r}{5cm}
  \centering
\includegraphics[width=4cm]{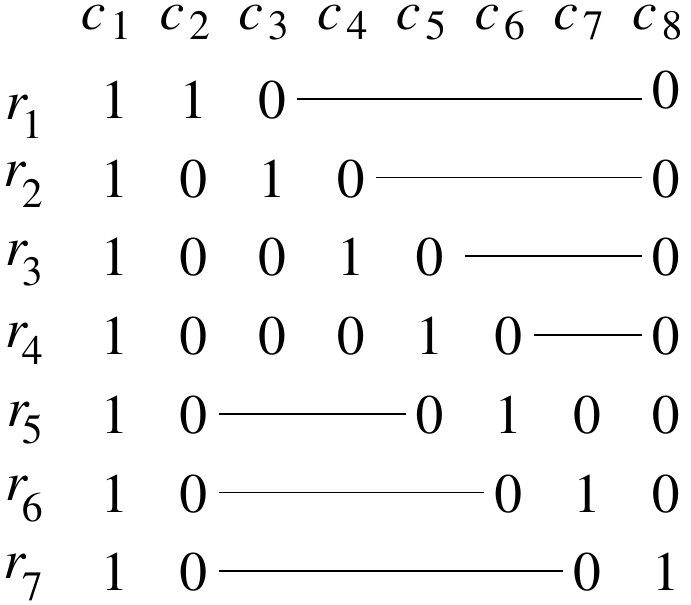}
\caption{A matrix not verifying the C1P and such
  that each set of 3 rows is a MCS.}
 \label{MCS-exp}
\end{wrapfigure}

\noindent
However, it is not difficult to build examples of  matrices such that the
number of MCS is polynomial or even exponential in the number of
rows.

 Figure \ref{MCS-exp} shows such an example in which each sub set of
$3$ rows is a MCS. Thus, such a construction with $m$ rows
gives $C^m_3$ $= O(m^3)$ MCS. Note that, on this example, a single
row is included in $O(m^2)$ MCS.

\begin{figure}[htb]
  \centering
\includegraphics[width=9cm]{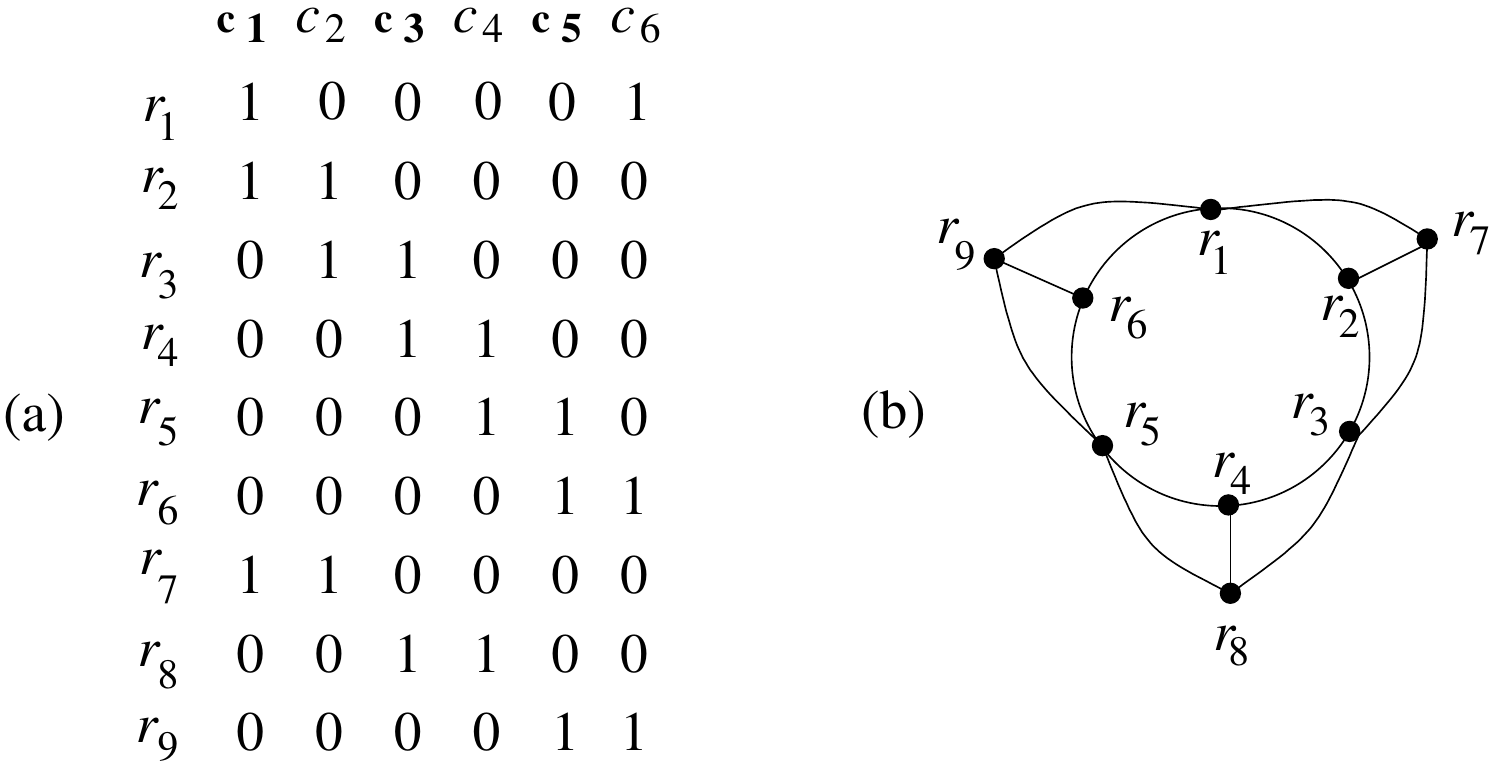}
\caption{(a) A matrix not verifying the C1P and such that the number 
of MCS is exponential in the  number of rows. (b) A row intersection
graph of the matrix whose vertices correspond to the rows of the matrix,
and such that there exists an edge between two rows $r_i$ and $r_j$ 
if $r_i\cap r_j \neq \emptyset$.}
 \label{MCS-exp2}
\end{figure}

 Figure \ref{MCS-exp2}-(a) shows another example where the number of MCS
 is exponential in the number of rows. Let $k$ be the number of nodes
 of {\em external} rows, which are $r_7,r_8,$ and $r_9$ on the
 figure. The total number of rows is $3k$, the number of columns $2k$,
 and the number of MCS is $2^k$ since any induced chordless cycle in
 the row intersection graph of the matrix  (Figure \ref{MCS-exp2}-(b)) 
constitutes a MCS.

From a computational point of view, the first question that arises is the
following: is a given row $r\in {\cal R}$ included in at least one MCS ? 
This question has been raised in \cite{BBCC2004}, recalled in 
\cite{CHSY2009,Chauve08} and recently solved in polynomial time
$O(m^6n^5(m+n)^2log(m+n))$ in \cite{Blin2011}. This currently fastest 
algorithm is based on the identification of minimal Tucker forbidden 
submatrices \cite{T1972,D2009}.

In this paper we present a new simpler $O(m^2n^2 + nm^7)$ time
algorithm for deciding if a given row belongs to at least one MCS and
if true exhibit one. Our algorithm is based on an
alternative approach considering minimal forbidden induced subgraphs
of interval graphs \cite{LB62} instead of Tucker matrices. Moreover,
our central paradigm consists in reducing the recognition of complex
forbidden induced subgraphs to the detection of induced cycles in
ad-hoc graphs, while in \cite{Blin2011} only induced paths are
considered. Our approach is faster and simpler, but a limit
shared by both approaches 
resides in  avoiding to report the number of MCS to which a given row belongs.

\section{MCS and Forbidden induced subgraphs}


The \emph{row-column intersection graph} of a 0-1 matrix 
$M=({\cal R}, {\cal C})$ is a vertex-colored 
bipartite graph $G_{RC}(M)$ whose set of vertices is ${\cal R}\cup {\cal C}$ ;  
the vertices corresponding to rows (resp. columns) are black (resp. white) ;
there exists an edge between two rows $r_i\in {\cal R}$ and $r_j\in {\cal R}$ 
if $r_i\cap r_j \neq \emptyset$, and there exists an edge between a row 
$r\in {\cal R}$ and a column $c\in {\cal C}$ if $c\in r$. 

It should be noted that 
a column vertex (white) is only connected to row vertices (black).

The \emph{neighborhood} $N(r)$ of a row $r$ is the set of rows intersecting $r$,
$N(r)= \{x\in {\cal R} ~:~ r\cap x \neq \emptyset\}$ and  
$N(r_i,r_j)= N(r_i)\cap N(r_j)$. The \emph{span} $L(c)$ of a column $c$ 
is the set of rows containing $x$, $L(c)= \{r\in {\cal R} ~:~ c\in r\}$.

\begin{figure}[h]
  \centering
\includegraphics[width=11cm]{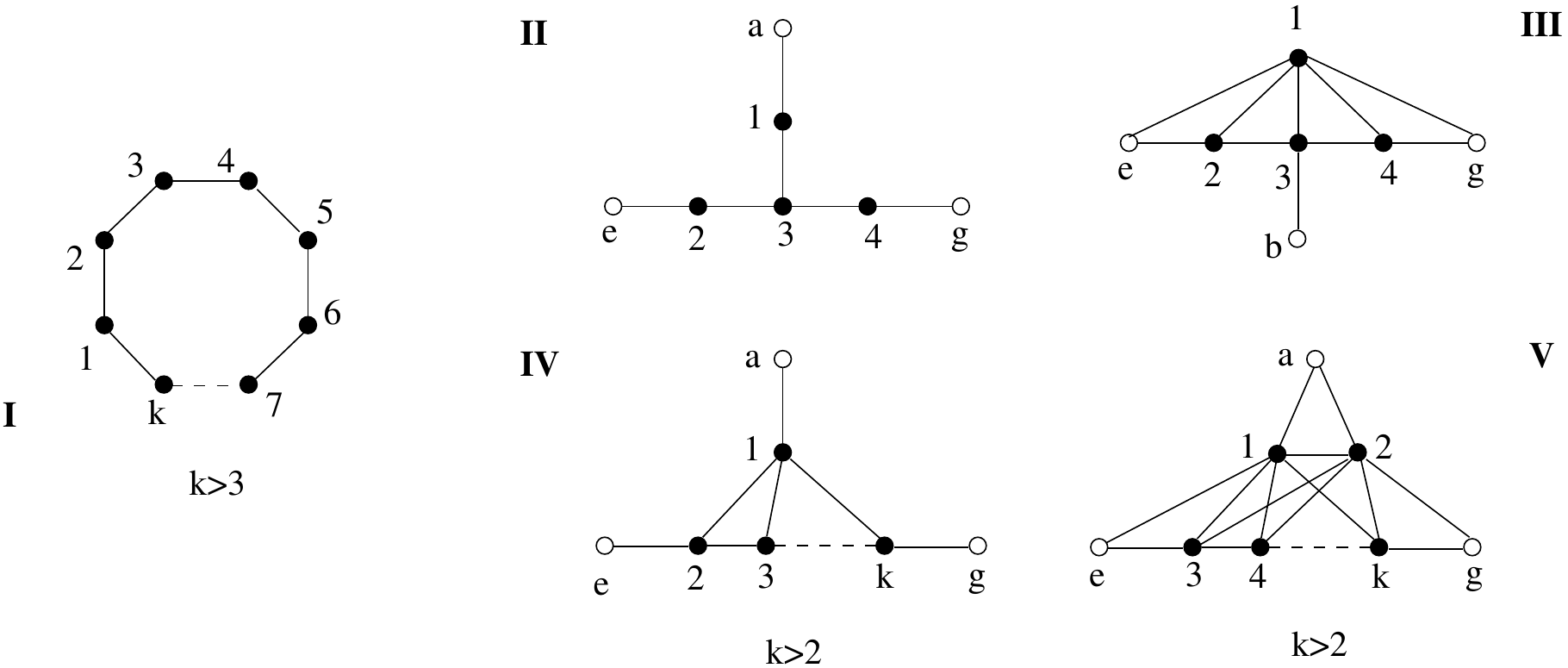}
\caption{Forbidden induced subgraphs for the row-column intersection graph of $M=({\cal R}, {\cal C})$ to verify C1P.}
 \label{forbid}
\end{figure}

\begin{theorem}[\cite{LB62}, Theorem 4]
\label{thm-fordid}
A 0-1 matrix $M=({\cal R}, {\cal C})$ verifies the C1P if and only 
if its row-column intersection 
graph does not contain a forbidden induced subgraph of the form I, II, III, IV, 
or V (Figure \ref{forbid}).
\end{theorem}

\begin{property}
From Theorem \ref{thm-fordid}, a set ${\cal S} \subseteq {\cal R}$ is a MCS
if the row-column intersection graph $G_{RC}({\cal S},{\cal C})$ contains a 
subgraph of the 
form I, II, III, IV, or V; and for any ${\cal T} \subset {\cal S}$, 
$G_{RC}({\cal T},{\cal C})$ does not contain a subgraph of the form I, II, III, 
IV, or V. 
\label{feat-mcs}
\end{property}

Given a MCS ${\cal S}\subseteq {\cal R}$, a forbidden induced subgraph 
contained in  
$G_{RC}({\cal S},{\cal C})$ is said to be \emph{responsible} for the MCS 
${\cal S}$. 
If this forbidden induced subgraph is of the form I (resp. II; III; IV; V), 
we simply say that ${\cal S}$ is a MCS of the form I (resp. II; III; IV; V).

\begin{definition}
A row of a MCS  ${\cal S}$ that intersects all other rows of ${\cal S}$
is called a \emph{kernel} of ${\cal S}$. In a forbidden  induced subgraph  
responsible for $ {\cal S}$, any kernel of ${\cal S}$ constitutes a  black 
vertex that is connected to all other  black vertices.
\label{def-kernel}
\end{definition}

\begin{property}
Note that an induced subgraph of the form II, III, IV, or V necessarily contains
at least one kernel, while an induced subgraph of the form I contains no kernel.
\label{mcs-kernel}
\end{property}

We denote by $G_{R}(M)$, the subgraph of $G_{RC}(M)$ induced by the set of rows 
${\cal R}$, thus containing only black vertices.

\noindent
{\bf Graph sizes.} $G_{R}(M)$ has $m$ vertices and at most $\min(mn,m^2)$ 
edges, while $G_{RC}(M)$  has $m+n$ vertices and at most $\min((m+n)^2,m^2n)$ 
edges.

\section{A global algorithm}

\begin{figure}[htb]
  \centering
\includegraphics[width=10cm]{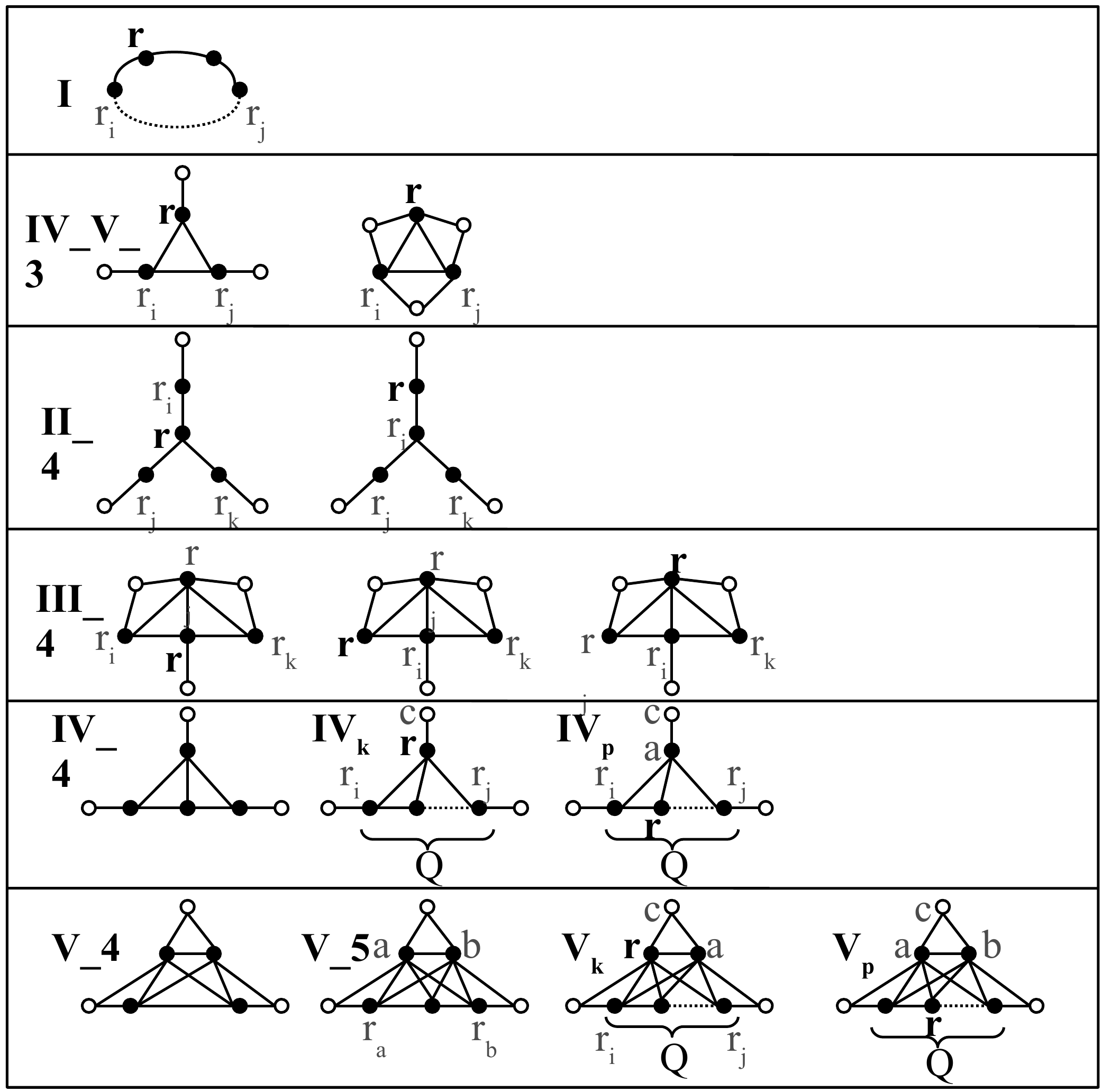}
\caption{The different steps of the algorithm: in each case, when row $r$ has 
a specific location in the forbidden induced subgraph that is looked for, this 
location is indicated in bold character. Other rows and columns of the 
forbidden induced subgraph are indicated in grey color characters.}
 \label{Algo}
\end{figure}

Our algorithm to decide if a row $r\in {\cal R}$ of a 0-1 matrix 
$M=({\cal R},{\cal C})$
belongs to at least one MCS, is based on a sequence of algorithms for finding 
a forbidden subgraph of $G_{RC}(M)$ responsible for a MCS 
containing $r$. It looks for forbidden subgraph of the form I, III, II, IV, V, 
in the following order: 1. MCS of type I, 2. MCS of size $3$ (types IV or V),
3. MCS of type II, 4. MCS of type III, 5. MCS of type IV and size larger or 
equal to $4$, and MCS of type V and size larger or equal to $4$. See Figure  
\ref{Algo}  for an overview. The steps 1 to 4 are based on straightforward 
brute-force algorithms, while the two last steps relies to a reduction to the 
detection of induced chordless cycles in ad-hoc graphs.

In the following, we simply write 
$G_{RC}(M)$ as $G$ and $G_{R}(M)$ as $G_R$.

\subsection{Step 1: Forbidden induced subgraph I}

We first test if $r$ belongs to a MCS  of the form I.
If it is true, then $r$ belongs to an induced chordless cycle of $G$
of length at least $4$ containing only black vertices. 
Such a cycle exists in $G$ if and only if is also a chordless
cycle in $G_{R}$ since $G_{R}$ is the subgraph of $G$ induced 
by the set of rows ${\cal R}$.
Thus it suffices to search for an induced chordless cycle 
in $G_{R}$.

\begin{algorithm}[htpb]   
\rule{11.7cm}{0.01cm}
\caption{Check\_I ($r$, $G_R$) -- $O(m^5)$}
\rule{11.7cm}{0.01cm}
\\               
{\bf Input:} a row $r$, the subgraph $G_R$.\\
{\bf Output:} returns a MCS ${\cal S}$ given by a forbidden induced subgraph of 
the form I containing $r$ if such a MCS exists, otherwise returns  "NO''.
\rule{11.7cm}{0.01cm}               
\begin{algorithmic}[1] 
\FOR{ any $P_4$ of $G_{R}$ containing $r$}
\STATE Consider the graph $G'$ obtained from $G_{R}$ after removing the two 
internal vertices of the $P_4$ and their neighborhood from the graph, and 
consider the extremities $r_i$ and $r_j$ of the $P_4$
\IF{ there exists a $r_ir_j$-path in $G'$}
\STATE find a chordless path $P$ in this graph linking $r_i$ and $r_j$.
\STATE return the set of vertices of the $P_4$ plus the set of vertices of $P$
\ENDIF
\ENDFOR
\STATE return "NO''
\end{algorithmic}
\rule{11.7cm}{0.01cm}
\end{algorithm}

\begin{proposition}
  Algorithm Check\_I is correct and runs in worst case $O(m^5)$ time.
\end{proposition}
\begin{proof}
The correctness of Algorithm \mbox{Check\_I} comes from the fact that,  
$r$ is contained in a MCS of the form I if and only if $r$ belongs to 
an induced chordless cycle of $G_{R}$ of length at least $4$ whose set
of vertices ${\cal S}$ constitutes the MCS (Figure \ref{Algo}.I). 
A $P_4$ of $G_{R}$ is an induced chordless path of $G_{R}$ containing
$4$ vertices.
In this case, Algorithm \mbox{Check\_I} returns such a set of vertices
since an induced chordless cycle of $G_{R}$ of length at least $4$
containing $r$ is a $P_4$ containing $r$ whose extremities are
linked by a chordless path in the subgraph of $G$ that does not contain
the neighborhood of the internal vertices of the $P_4$. This set ${\cal
  S}$ cannot contain a smaller subset of rows that is a MCS, as no
subset of $S$ can be a MCS of the form I, or a MCS of any other form
because of Property \ref{mcs-kernel}.

Algorithm \mbox{Check\_I} might be implemented in $O(m^5)$. 
The test performed on a give $P_4$ containing $r$ (lines 2-5 of the algorithm)
can be achieved in  $O(\min(mn,m^2)+m\log m)$ as follows: 
removing the neighborhood of its internal vertices
might be done in $\min (mn,m^2)$ time, and finding a chordless path
between the two extremities might be performed using Dijkstra's
algorithm 
in $O(\min(mn,m^2)+m\log m)$ time.
Enumerating all $P_4$ containing $r$ might be done in time $O(m^3)$ using
a BFS from $r$ stopping at depth $4$.
Eventually, the whole algorithm is in  $O(m^3 (\min(mn,m^2)+m\log m)) = O(m^5)$ time.
\end{proof}

\noindent
{\bf Precomputation.} In the following steps, we assume that the following 
precomputations have been achieved:
\begin{itemize}
\item For any triplet of rows $(r,r_i,r_j)$ that are pairwise intersecting,
\emph{i.e} each couple is an edge in $G$, 
$r-(r_i\cup r_j)$ and $(r_i\cap r_j)-r$ are precomputed ; 
\item Two rows  $r_i$ and $r_j$ are  \emph{overlapping} if  
$r_i\cap r_j \neq\emptyset$ and $r_i - r_j\neq\emptyset$ and 
$r_j - r_i\neq\emptyset$.
The overlapping relation between any couple of rows is precomputed ; 
\item For any quadruplet of rows $(r, r_i,r_j,r_k)$ such that $r_i,r_j$, and 
$r_k$ overlap $r$,  $r - (r_i\cap r_j\cap r_k)$ is precomputed.
\end{itemize}

All those precomputations can simply be performed in $O(m^4n)$ time
using straightforward algorithms, that is, scanning the $n$ columns of
the input matrix for each triplet or quadruplet of rows.  

\subsection{Step 2: Forbidden induced subgraph responsible for a MCS of size $3$}

We test here if $r$ belongs to a MCS of size $3$. A MCS of size $3$ is 
necessarily caused by a forbidden induced subgraph of the form IV or V.
As a consequence, the following property is immediate.

\begin{property}
A MCS of size $3$ is always composed of $3$ rows that are pairwise 
overlapping.
\label{mcs-3}
\end{property}

\begin{algorithm}[htpb]                    
\rule{11.7cm}{0.01cm}
\caption{Check\_IV\_V\_3 ($r$, $G$) -- $O(m^2)$}
\rule{11.7cm}{0.01cm}
\\               
{\bf Input:} a row $r$, the row-column intersection graph $G$.\\
{\bf Output:} returns a MCS ${\cal S}$ of size $3$ given by a forbidden 
induced subgraph of the form IV or V containing $r$ if such a MCS exists, 
otherwise returns  "NO''.
\rule{11.7cm}{0.01cm}              
\begin{algorithmic}[1] 
\FOR{ any couple $(r_i,r_j)$ of black vertices that both overlap $r$, and overlap each other}
\IF{ $r-(r_i\cup r_j)\neq \emptyset$ and $r_i-(r\cup r_j)\neq \emptyset$
and $r_j-(r\cup r_i)\neq \emptyset$}
\STATE return $\{r,r_i,r_j\}$
\ENDIF
\IF{ $(r_i\cap r_j)-r\neq \emptyset$ and $(r\cap r_j)-r_i\neq \emptyset$
and $(r\cap r_i)-r_j\neq \emptyset$}
\STATE return $\{r,r_i,r_j\}$
\ENDIF
\ENDFOR
\STATE return "NO''
\end{algorithmic}
\rule{11.7cm}{0.01cm}
\end{algorithm}

\begin{proposition}
Algorithm \mbox{Check\_IV\_V\_3} is correct and runs in $O(m^2)$ time.
\end{proposition}
\begin{proof}
The correctness of Algorithm \mbox{Check\_IV\_V\_3} comes from the fact that, 
$r$ is contained in a MCS of size $3$ if and only if this MCS is caused 
by a forbidden induced subgraph of the form IV or V (Property \ref{mcs-3}). 
Thus, $r$ should belong to a triplet of rows $(r,r_i,r_j)$ that are pairwise
overlapping, and satisfy the conditions given in:
\begin{itemize}
\item either, line 2 of the algorithm to produce a forbidden induced subgraph of the form IV (left-end graph in Figure \ref{Algo}.IV\_V\_3),
\item or, line 5 of the algorithm to produce a forbidden induced subgraph of the form V (right-end graph in Figure \ref{Algo}.IV\_V\_3).
\end{itemize}
In both cases, Algorithm \mbox{Check\_IV\_V\_3} returns the set $\{r,r_i,r_j\}$ 
as a MCS if such a set of rows exists. This set cannot contain a smaller subset
of rows that is a MCS as $3$ is the minimum size of any MCS.

Algorithm Check\_IV\_V\_3 runs in $O(m^2)$ time since, given $r$,
there might be $O(m^2)$ couples $(r_i,r_j)$ on which the tests performed
(lines 2-8 of the algorithm) might be
achieved in $O(1)$, thanks to the precomputations that have been done.
\end{proof}

\subsection{Step 3: Forbidden induced subgraph II}

We test here if $r$ belongs to a MCS of the form II, with
the assumption that $r$ is not contained in any MCS of size $3$.
Note that such a MCS is of size $4$.

\begin{algorithm}[htpb]                    
\rule{11.7cm}{0.01cm}
\caption{Check\_II\_4 ($r$, $G$)--$O(m^3)$}
\rule{11.7cm}{0.01cm}
\\               
{\bf Input:} a row $r$, the row-column intersection graph $G$.\\
{\bf Assumption:} $r$ is not contained in a MCS of size $3$.\\
{\bf Output:} returns a MCS ${\cal S}$ given by a forbidden induced subgraph of 
the form II containing $r$ if such a MCS exists, otherwise returns  "NO''.
\rule{11.7cm}{0.01cm}
\begin{algorithmic}[1]
\FOR{ any triplet $(r_i,r_j,r_k) $ of black vertices such that $r_i,r_j,r_k$ overlap $r$}
\IF{ there are no edges $(r_i,r_j)$, $(r_i,r_k)$, and $(r_j,r_k)$ in $G$}
\STATE return $\{r,r_i,r_j,r_k\}$
\ENDIF
\ENDFOR 
\FOR{ any triplet $(r_i,r_j,r_k) $ of black vertices of such that $r_i$ overlaps $r$, and $r_j,r_k$ overlap $r_i$}
\IF{ there are no edges $(r,r_j)$, $(r,r_k)$, or $(r_j,r_k)$ in $G$}
\STATE return $\{r_i,r,r_j,r_k\}$
\ENDIF
\ENDFOR 
\STATE return "NO''
\end{algorithmic}
\rule{11.7cm}{0.01cm}
\end{algorithm}

\begin{proposition}
Algorithm \mbox{Check\_II\_4} is correct and  runs in $O(m^3)$ time.
\end{proposition}
\begin{proof}
The correctness of Algorithm \mbox{Check\_II\_4} comes from the fact that,  if $r$ belongs to a MCS of the form II,
then $r$ should belong to a quadruplet of rows $(r,r_i,r_j,r_k)$ such that one
these rows is a kernel, and the three other rows do not intersect each other. 
Thus, the row $r$ is:
\begin{itemize}
\item either, a kernel of the MCS, tested in lines 1-5 of the algorithm (left-end graph 
in Figure \ref{Algo}.II\_4),
\item or, not a kernel of the MCS tested in lines 6-10 of the algorithm (right-end 
graph in Figure \ref{Algo}.II\_4).
\end{itemize}
In both cases, Algorithm \mbox{Check\_II\_4} returns the set 
$\{r,r_i,r_j,r_k\}$ as a MCS if such a set of rows exists. 
This set cannot contain a smaller subset of rows that is a MCS as this 
subset would be a subset of $3$ rows that cannot satisfy Property \ref{mcs-3}.

Algorithm Check\_IV\_V\_4 runs in $O(m^3)$ time since all the tests
performed on a given triplet $(r_i,r_j,r_k)$ in lines 2-4 and 7-9 of 
algorithm can be achieved in $O(1)$, and given $r$ there might be 
$O(m^3)$ such triplets.
\end{proof}

\subsection{Step 4: Forbidden induced subgraph III}

We test here if $r$ belongs to a MCS of the form III, 
with the assumption that  $r$ is not contained in a MCS of size $3$.
Note that such a MCS is of size $4$.

\begin{algorithm}[htpb]                    
\rule{11.7cm}{0.01cm}
\caption{Check\_III\_4 ($r$, $G$)--$O(m^3)$}
\rule{11.7cm}{0.01cm}
\\
{\bf Input:} a row $r$, the row-column intersection graph $G$.\\
{\bf Assumption:}  $r$ is not contained in a MCS of size $3$.\\
{\bf Output:} returns a MCS ${\cal S}$ given by a forbidden induced subgraph of 
the form III containing $r$ if such a MCS exists, otherwise returns  "NO''.
\rule{11.7cm}{0.01cm}
\begin{algorithmic}[1]
\FOR{ any triplet $(r_i,r_j,r_k) $ of black vertices such that $r_i,r_j,r_k$ overlap $r$, and $r - (r_i\cap r_j\cap r_k) \neq \emptyset$}
\IF{ there are no edge $(r_i,r_k)$ in $G$, and $(r_i\cap r_j)-r \neq \emptyset$, and $(r_j\cap r_k)-r \neq \emptyset$}
\STATE return $\{r,r_i,r_j,r_k\}$
\ENDIF
\ENDFOR 

\FOR{ any triplet $(r_i,r_j,r_k) $ of black vertices of such that  $r_i$ overlaps $r$, and $r_j,r_k$ overlap $r_i$, and  $r_i - (r\cap r_j\cap r_k) \neq \emptyset$, and $\{r_i,r_j,r_k\}$ is not a MCS}
\IF{ there are no edge $(r,r_k)$ in $G$, and $(r\cap r_j)-r_i \neq \emptyset$, and $(r_j\cap r_k)-r_i \neq \emptyset$ }
\STATE return $\{r_i,r,r_j,r_k\}$
\ENDIF
\ENDFOR 
\FOR{ any triplet $(r_i,r_j,r_k) $ of black vertices of such that $r_i$ overlaps $r$, and $r_j,r_k$ overlap $r_i$, and  $r_i - (r\cap r_j\cap r_k) \neq \emptyset$}
\IF{ there are no edge $(r_j,r_k)$ in $G$, and $(r_j\cap r)-r_i \neq \emptyset$, and $(r\cap r_k)-r_i \neq \emptyset$}
\STATE return $\{r_i,r_j,r,r_k\}$
\ENDIF
\ENDFOR 
\STATE return "NO''
\end{algorithmic}
\rule{11.7cm}{0.01cm}
\end{algorithm}

\begin{proposition}
Algorithm \mbox{Check\_III\_4} is correct and runs in $O(m^3)$ time.
\end{proposition}
\begin{proof}
The correctness of Algorithm \mbox{Check\_III\_4} comes from the fact that, 
$r$ belongs to a MCS of the form 
III if and only if $r$ should belong to a quadruplet of rows 
$(r,r_i,r_j,r_k)$ included in an induced subgraph of the form III
such that two of these rows are kernels of the subgraph, and one of these 
kernels contains a column of the induced subgraph that is not shared 
with any of the other rows. Let us call this kernel kernel\_1, and the 
other kernel kernel\_2. For example in the left-end graph 
in Figure \ref{Algo}.III\_4, kernel\_1=$r$, and kernel\_2=$r_j$.

Thus, the row $r$ is:
\begin{itemize}
\item either, kernel\_1, tested in lines 1-5 of the algorithm (left-end graph 
in Figure \ref{Algo}.III\_4),
\item or, not a kernel, tested in  lines 6-10 of the algorithm (middle 
graph in Figure \ref{Algo}.III\_4).
\item or, kernel\_2, tested in  lines 11-15 of the algorithm (right-end 
graph in Figure \ref{Algo}.III\_4).
\end{itemize}
In the first, and third cases, the set  $\{r_i,r_j,r_k\}$ cannot be a MCS
because such a set cannot satisfy Property \ref{mcs-3}
In all cases, Algorithm \mbox{Check\_III\_4} returns the set 
$\{r,r_i,r_j,r_k\}$ as a MCS if 
such a set of rows exists, and $\{r_i,r_j,r_k\}$ is not a MCS 
(in the second case). 
Since we made the assumption that $r$ is not contained in a MCS of size $3$,
there cannot exists a smaller subset of  $\{r_i,r_j,r_k\}$ containing $r$
that is a MCS.

Algorithm Check\_III\_4 runs in $O(m^3)$ time using a similar proof
as the complexity proof for Check\_IV\_V\_4: all the tests performed by 
the algorithm (lines 2-4, 7-9, and 12-14 of the algoritms) on a given triplet 
$(r_i,r_j,r_k)$ are achieved in $O(1)$ thanks to the precomputations, 
and given $r$ there might be $O(m^3)$ such triplets.
\end{proof}

\subsection{Step 5: Forbidden induced subgraph IV}

We test here if $r$ belongs to a MCS of the form IV, with the assumption that  
$r$ is contained, neither in a MCS of size $3$, nor in a MCS of type I.
Depending on whether the size of the MCS is $4$ or larger than $4$, 
we describe two algorithms.

\subsubsection{MCS of size $4$}

We first test if $r$ belongs to a MCS of the form IV of size $4$.
We look for a triplet of rows $(r_i,r_j,r_k)$ such that 
the set $\{r,r_i,r_j,r_k\}$ is a MCS of the form IV (Figure \ref{Algo}.IV\_4).
In an induced subgraph of the form IV containing $4$ rows  $\{r,r_i,r_j,r_k\}$, 
two rows are kernels, and in that case, $r$ is either a kernel of the MCS, 
or not. If $r$ is a kernel, then it is either a kernel --called kernel\_1-- 
containing a column of the induced subgraph that is not shared with any 
of the other rows , or not --called kernel\_2--. For example, in the 
left-end graph in Figure 4.IV\_4, the two kernel are the two central black
vertices of the graph: the top one is a kernel\_1, and the bootom one a
 kernel\_2.
Algorithm \mbox{Check\_IV\_4} looks for each of these configurations: 
\begin{itemize}
\item  $r$ is a kernel\_1, tested in lines 1-5 of the algorithm;  
\item $r$ is not a kernel,tested in lines 6-10 of the algorithm; 
\item $r$ is a kernel\_2, tested in lines 11-15 of the algorithm.
\end{itemize}
The proof of the correctness of Algorithm \mbox{Check\_IV\_4} is similar to the
proof for Algorithm \mbox{Check\_III\_4}.

\begin{algorithm}[htpb]                      
\rule{11.7cm}{0.01cm}

\caption{Check\_IV\_4 ($r$ , $G$) -- $O(m^3)$} 
\rule{11.7cm}{0.01cm}
\\
{\bf Input:} a row $r$, the row-column intersection graph $G$.\\
{\bf Assumption:}  $r$ is not contained in a MCS of size $3$.\\
{\bf Output:} returns a MCS ${\cal S}$ of size $4$ given by a forbidden 
induced subgraph of the form IV containing $r$ if such a MCS exists, otherwise returns  "NO''.
\rule{11.7cm}{0.01cm}

\begin{algorithmic}[1]
\FOR{ any triplet $(r_i,r_j,r_k) $ of black vertices such that $r_i,r_j,r_k$ 
are connected to $r$, and $r - (r_i\cap r_j\cap r_k) \neq \emptyset$}
\IF{ there are no edge $(r_i,r_k)$ in $G$, and $(r_i\cap r_j) \neq \emptyset$, and $(r_j\cap r_k) \neq \emptyset$, and $r_i - (r \cup r_j) \neq \emptyset$, and
$r_k - (r \cup r_j) \neq \emptyset$}
\STATE return $\{r,r_i,r_j,r_k\}$
\ENDIF
\ENDFOR 

\FOR{ any triplet $(r_i,r_j,r_k) $ of black vertices of such that $r_i$ 
is connected to $r$, and $r_j,r_k$ are connected to $r_i$, and  $r_i - (r\cap r_j\cap r_k) \neq \emptyset$, and $\{r_i,r_j,r_k\}$ is not a MCS}
\IF{ there are no edge $(r,r_k)$ in $G$, and $(r\cap r_j) \neq \emptyset$, and $(r_j\cap r_k) \neq \emptyset$, and $r - (r_i \cup r_j) \neq \emptyset$, and
$r_k - (r_i \cup r_j) \neq \emptyset$}
\STATE return $\{r_i,r,r_j,r_k\}$
\ENDIF
\ENDFOR 
\FOR{ any triplet $(r_i,r_j,r_k) $ of black vertices of such that $r_i$ 
is connected to $r$, and $r_j,r_k$ are connected to $r_i$, and  $r_i - (r\cap r_j\cap r_k) \neq \emptyset$}
\IF{ there are no edge $(r_j,r_k)$ in $G$, and $(r_j\cap r) \neq \emptyset$, and $(r\cap r_k) \neq \emptyset$, and $r_j - (r \cup r_i) \neq \emptyset$, and
$r_k - (r \cup r_i) \neq \emptyset$}
\STATE return $\{r_i,r_j,r,r_k\}$
\ENDIF
\ENDFOR 
\STATE return "NO''

\end{algorithmic}
\rule{11.7cm}{0.01cm}
\end{algorithm}

\begin{proposition}
Algorithm $\mbox{Check\_IV\_4}$ is correct and runs in $O(m^3)$ time.
\end{proposition}
\begin{proof}
The proof for Algorithm \mbox{Check\_IV\_4} is similar to the
proof for Algorithm \mbox{Check\_III\_4}.
\end{proof}

\subsubsection{MCS of size larger than $4$}

We test here if $r$ belongs to a MCS of the form IV of size larger than $4$.
A MCS of the form IV of size larger than $4$ contains one and only
one kernel. Depending on whether $r$ is the kernel or not, we
distinguish two cases here.\\

{\bf Case 1: If row $r$ is the kernel of the MCS}\\

Algorithm \mbox{Check\_IV$_k$} recovers a MCS ${\cal S}$ of the form IV of size 
larger than $4$ containing $r$ as a kernel, with the assumption that  $r$ 
is not contained in a MCS of size $3$ (Figure \ref{Algo}.IV$_k$).
The principle of the algorithm relies in first choosing the column 
$c\in {\cal C}$, of the
forbidden induced subgraph of type IV responsible for ${\cal S}$, that is 
contained in $r$, and in no other row of the MCS  
(see Figure \ref{Algo}.IV$_k$).
Next, it considers the subgraph $H$ of $G$ induced by the set of black 
vertices (rows) that are neighbors of $r$, but do not contain the column $c$.
We denote this subgraph by  $H = G[N(r)-L(c)]$.
Then, it looks for a set of rows $Q$, constituting a chordless path in $H$, 
such that $\{r\}\cup Q$ is a MCS of the form IV.

\begin{algorithm}[htpb]                      
\rule{11.7cm}{0.01cm}

\caption{Check\_IV$_k$ ($r$, $G$) -- $O(nm^2)$} 
\rule{11.7cm}{0.01cm}
\\
{\bf Input:} a row $r$, the row-column intersection graph $G$.\\
{\bf Assumption:}  $r$ is not contained in a MCS of size $3$.\\
{\bf Output:} returns a MCS ${\cal S}$ of size larger that $4$ given by a 
forbidden induced subgraph of the form IV whose kernel is $r$ if such a 
MCS exists, otherwise returns  "NO''.
\rule{11.7cm}{0.01cm}

\begin{algorithmic}[1]
\FOR{ any column $c \in r$}
\STATE $H = G[N(r)-L(c)]$
\FOR{any connected component $C$ of $H$}
\STATE pick a a couple $(r_i,r_j)$ of black vertices in $C $ that satisfies 
1) $r_i$ and $r_j$ are not connected, and 2) $r_i,r_j$  overlap $r$.
\STATE find a chordless path $P$ in $C$ linking $r_i$ and $r_j$
\STATE pick the smallest subpath $Q$ of $P$ linking two vertices $r'_i$ and 
$r'_j$, such that the couple $(r'_i,r'_j)$ also satisfies 1) and 2)
\STATE return $\{r\}\cup Q$
\ENDFOR
\ENDFOR
\STATE return ``NO''
\end{algorithmic}
\rule{11.7cm}{0.01cm}
\end{algorithm}

\begin{proposition}
Algorithm $\mbox{Check\_IV}_k$ is correct and runs in $O(nm^2)$ time.
\end{proposition}
\begin{proof}
 Note that, if the MCS exists, then all the rows belonging to the MCS, 
except $r$, belong to a same 
connected component of $H$. Thus, in each connected component of $H$, the
algorithm looks for a chordless path $Q$ linking two vertices $r_i,r_j$ 
satisfying 1) $r_i$ and $r_j$ are not connected, and 2) $r_i,r_j$  overlap 
$r$, and 3) $Q$ does not contain any smaller subpath satisfying conditions 
1) and 2). 
These conditions are necessary and sufficient for the set  $\{r\}\cup Q$
to form the rows of a induced subgraph of the form $IV$. The set  $\{r\}\cup Q$
cannot contain a subset that is a MCS as such a smaller MCS should be:
\begin{itemize}
 \item either a MCS of size $3$ including $r$, which impossible by assumption, 
 \item or a MCS of type II or III necessarily including $r$ as kernel,
 \item or a MCS of  type IV and size larger than $3$  having $r$ as kernel.
\end{itemize}
The two last cases are also impossible, since $Q$ would not have satisfy
condition 3) in these cases.

Next, there might be $n$ columns $c\in r$ and up to $m^2$
couples $(r_i,r_j)$ of black vertices to test before 
finding a valid couple $(r_i,r_j)$ satisfying
the conditions in line 4 of the algorithm. 
Up to this point, the complexity is in $O(nm^2)$. Assume now
that such a couple exist. Then finding a chordless path between $r_i$
and $r_j$ might be done by searching for a shortest path between $r_i$
and $r_j$ in the connected component $C$ using Dijkstra's algorithm, 
which thus requires at worst $O(min(mn,m^2)+m\log m)$ time. The path 
is of length at most $m$, and thus identifying $r'_i$ and $r'_j$ is 
bounded by testing each pair on this path in $C$, which requires at 
worst $O(m^2)$
time. Thus, in total, the algorithm is $O(nm^2)$ worst case time.
\end{proof}

{\bf  Case 2: If row $r$ is not the kernel of the  MCS}\\

Algorithm \mbox{Check\_IV$_p$} recovers a MCS ${\cal S}$ of the form IV 
of size larger 
than $4$ containing $r$, but not as a kernel, with the assumptions that $r$ is 
not contained in a MCS of size $3$, and $r$ does not belong to an induced 
chordless cycle of $G_R$ (Figure 4.IV$_p$). The principle of the algorithm
consists in first choosing the kernel $a$ of  ${\cal S}$ among the black 
vertices (rows) neighbors of $r$, and the column $c\in {\cal C}$, of the
induced subgraph of type IV responsible for ${\cal S}$, that is contained 
in $a$, but in no other row of the MCS.
(see Figure \ref{Algo}.IV$_p$).
Next, the algorithm calls Algorithm \mbox{Check\_IV} to look for the MCS 
${\cal S}$ with $r$, $a$, $c$, and $G$ given as parameters.

\begin{algorithm}[htpb]                      
\rule{11.7cm}{0.01cm}
\caption{Check\_IV$_p$ ($r$, $G$) -- $O(nm^6)$} 
\rule{11.7cm}{0.01cm}
\\
{\bf Input:} a row $r$, the row-column intersection graph $G$.\\
{\bf Assumption:}  $r$ is not contained in a MCS of size $3$.\\
$~~~~~~~~~~~~~~~~~~~~~r$ does not belong to an induced chordless cycle of $G_R$.\\
{\bf Output:} returns a MCS ${\cal S}$ of size larger that $4$ given by a 
forbidden induced subgraph of the form IV containing $r$ whose kernel is 
not $r$ if such a MCS exists, otherwise returns  "NO''.
\rule{11.7cm}{0.01cm}
\begin{algorithmic}[1]                
\FOR{ any black vertex $a\in N(r)$}
\FOR{ any column $c\in a-r$}
\STATE return Check\_IV($r$, $a$, $c$, $G$)
\ENDFOR
\ENDFOR
\STATE return ``NO''
\end{algorithmic}
\rule{11.7cm}{0.01cm}
\end{algorithm}

Algorithm \mbox{Check\_IV} is called in Algorithm \mbox{Check\_IV$_p$}. It 
recovers a  MCS ${\cal S}$ of the form IV of size larger 
than $4$ containing $r$, given the row $r$, the  kernel $a$ of 
the MCS ${\cal S}$, and the column $c\in {\cal C}$, of the
induced subgraph of type IV responsible for ${\cal S}$, that is contained 
in $a$, but in no other row of the MCS (Figure \ref{Algo}.IV$_p$).

 \begin{algorithm}[htpb]                      
 \rule{11.7cm}{0.01cm}
 \caption{Check\_IV ($r$, $a$, $c$, $G$)-- $O(m^5)$}
 \rule{11.7cm}{0.01cm}
 \\
 {\bf Input:} two rows $r$ and $a$, and a column $c\in a$ such that
  $r \in (N(a)-L(c))$ .\\
 {\bf Assumption:}  $r$ is not contained in a MCS of size $3$.\\
 $~~~~~~~~~~~~~~~~~~~~~r$ does not belong to an induced chordless cycle of $G_R$.\\
 {\bf Output:} returns a MCS ${\cal S}$ of size larger that $4$ given by a 
 forbidden induced subgraph of the form IV containing $r$ and $a$, whose 
 kernel is $a$ if such a MCS exists, otherwise returns  "NO''.
 \rule{11.7cm}{0.01cm}
 \begin{algorithmic}[1]                
 \STATE $H = G[N(a)-L(c)]$
 \STATE let $C=(V_C,E_C)$ be the connected component of $H$ to which $r$ belongs.

 \STATE let $V_a$ be the set of vertices $V_a= \{u\in V_C ~:~ u-a \neq \emptyset\}$.
 \STATE let $E_a$ be the set of edges $E_a= \{(u,v)\in V_a^2 ~:~  u\cap v = \emptyset\}$.

 \STATE let $D=(V_D,E_D)$ be the graph such that $V_D = V_C$ and $E_D=E_C\cup E_a$.

 \STATE $Q$ =  Check\_I ($r$, $D_R$)
 \IF{$Q\neq$ "NO''}
 \STATE return $\{a\}\cup Q$
 \ENDIF
 \STATE return "NO''
 \end{algorithmic}
 \rule{11.7cm}{0.01cm}
 \end{algorithm}

\begin{proposition}
Algorithm $\mbox{Check\_IV}_p$ is correct, and runs in $O(nm^6)$ time.
\end{proposition}
\begin{proof}

The correctness and the complexity of $\mbox{Check\_IV}_p$ follows
directly from the the correctness and the complexity of 
Algorithm $\mbox{Check\_IV}$ that is called in Algorithm $\mbox{Check\_IV}_p$.

The correctness of $\mbox{Check\_IV}$ comes from the fact that, 
$r$ does not belong to any chordless cycle in the graph $C$ 
computed at line 2 of the algorithm by assumption.  
Then at line 6 of the algorithm, any
chordless cycle in the graph $D$ containing vertex $r$ necessarily
contains at least one edge $(r_i,r_j)$ belonging to the set
$E_a$. The number of edges belonging to the
set $E_a$ in such a chordless cycle $Q$ cannot be greater than $1$ 
as any couple of such edges in the chordless cycle would induce a chord. 
Indeed, if $Q$ contains more
than one edge belonging to $E_a$, any two such edges would have to
extremities in $V_a$, one from each of the two edges, that are not
connected in the graph $C$. These extremities would thus be linked 
by an edge in $E_a$, creating a chord for the cycle $Q$ in the graph $D$.

Therefore, the set of vertices of the chordless cycle $Q$ induces 
  a chordless path in $G$ such that each vertex of $Q$ is connected to 
 vertex $a$ by definition of the graph $H$, and the extremities
 $r_i$ and $r_j$ of $Q$ satisfy 1) $r_i$ and $r_j$ are not connected in $G$, 
 and 2) $r_i,r_j$  overlap $r$, and 3) $Q$ does not contain any smaller 
 subpath satisfying conditions 1) and 2). These conditions are necessary 
 and sufficient for the set  $\{a\}\cup Q$
 to form the rows of an induced subgraph of the form $IV$, and this set  
 cannot contain a smaller MCS since such a  MCS would be:
\begin{itemize}
 \item either a MCS of size $3$ including $a$, 
 \item or a MCS of type II or III necessarily including $a$ as kernel,
 \item or a MCS of  type IV and size larger than $3$  having $a$ as kernel.
\end{itemize}
The 3 cases are impossible,  since they would induce a chord from the
set $E_a$ in the chordless cycle induced by $Q$ in the graph $D$.

Algorithm $\mbox{Check\_IV}$ calls Algorithm $\mbox{Check\_I}$. 
Both algorithms have the same time complexity in $O(m^5)$ time.
It follows immediately that  Algorithm $\mbox{Check\_IV}_p$
runs  in $Onm^6$ time.
\end{proof}
 
\subsection{Step 6: Forbidden induced subgraph V}
 
We test here if $r$ belongs to a MCS of the form V, with the assumption that  
$r$ is contained neither in a MCS of size $3$, nor in a MCS of type I.
Depending on whether the size of the MCS is $4$, $5$ or larger than $5$, 
we describe three algorithms.

\subsubsection{MCS of size $4$ or $5$}

We first test if $r$ belongs to a MCS of the form V of size $4$ or $5$.
For a MCS of size 4, we look  for a triplet of rows 
$(r_i,r_j,r_k)$ such that the set $(r,r_i,r_j,r_k)$ is a MCS of the
form V. In such a case, we look for an induced
subgraph responsible for the MCS, containing $r,r_i,r_j,r_k$ as four black 
vertices pairwise connectedr, and we can pick three different couples of
$r,r_i,r_j,r_k$ such that each couple shares a column (white vertex) that 
is not shared with the two other of the MCS (see Figure \ref{Algo}.V\_4). 

\begin{algorithm}[htpb]                      
\rule{11.7cm}{0.01cm}
\caption{Check\_V\_4 ($r$ , $G$) -- $O(m^3)$} 
\rule{11.7cm}{0.01cm}
\\
{\bf Input:} a row $r$, the row-column intersection graph $G$.\\
{\bf Assumption:}  $r$ is not contained in a MCS of size $3$.\\
{\bf Output:} returns a MCS ${\cal S}$ of size $4$ given by a forbidden 
induced subgraph of the form V containing $r$ if such a MCS exists, otherwise returns  "NO''.
\rule{11.7cm}{0.01cm}
\begin{algorithmic}[1]
\FOR{ any triplet $(r_i,r_j,r_k) $ of black vertices such that $r_i,r_j,r_k$ 
are connected to  $r$, and are pairwise connected}
\IF{ $(r_i,r_j,r_k) $  is not a MCS, and $(r \cap r_i) - (r_j \cup r_k) \neq \emptyset$, and $(r_j \cap r_k) - (r \cup r_i) \neq \emptyset$}
\IF{$(r_i \cap r_j) - (r \cup r_k) \neq \emptyset$}
\STATE return $\{r,r_i,r_j,r_k\}$
\ENDIF
\IF{$(r \cap r_j) - (r_i \cup r_k) \neq \emptyset$}
\STATE return $\{r_i,r,r_j,r_k\}$
\ENDIF
\ENDIF
\ENDFOR 
\STATE return "NO''
\end{algorithmic}
\rule{11.7cm}{0.01cm}
\end{algorithm}

\begin{proposition}
Algorithm $\mbox{Check\_V\_4}$ is correct and runs in $O(m^3)$ time.
\end{proposition}
\begin{proof}
Algorithm \mbox{Check\_V\_4} looks for an induced subgraph with 
$4$ black vertices $\{r,r_i,r_j,r_k\}$, that are  pairwise connected 
to each other. 
These $4$ black vertices should be such that there exist three different 
couples of vertices among them, such that two couples are disjoint and the 
third one (called couple\_kernel) overlaps the two first, and the $2$ rows 
of each of these couples share a column that is not shared with the two 
other rows of the set. 
In this case, if $\{r_i,r_j,r_k\}$ is not a MCS, then the subgraph induced
by $\{r,r_i,r_j,r_k\}$ and the $3$ columns (white vertices) connected to 
the  $3$ couples
of rows is of the form V, and is responsible for a MCS $\{r,r_i,r_j,r_k\}$.
Algorithm \mbox{Check\_V\_4} looks for two cases, depending on whether
$r$ belong to couple\_kernel (lines 3-5), or not (lines 6-8). 

Next, all the tests performed by 
Algorithm \mbox{Check\_V\_4} (lines 2-9 of the algoritm) on a given triplet 
$(r_i,r_j,r_k)$ are achieved in $O(1)$ thanks to the precomputations, 
and given $r$ there might be $O(m^3)$ such triplets.
Thus, Algorithm \mbox{Check\_V\_4} runs in $O(m^3)$ time.
\end{proof}

Next, for a MCS of size 5, we look  for a quadruplet of rows 
$(r_i,r_j,r_k,r_l)$ such that the set $\{r,r_i,r_j,r_k,r_l\}$ is a MCS of the
form V (Figure \ref{Algo}.V\_5). Algorithm Check\_V\_5 looks for an
induced subgraph of the form V, consisting of $5$ rows (black vertices) 
$r,r_i,r_j,r_k,r_l$
that are pairwise connected, except for a on missing edge,
say $(r_a,r_b)$  in $\{r,r_i,r_j,r_k,r_l\}\times\{r,r_i,r_j,r_k,r_l\}$,
and three columns (white vertices) satisfying the configuration
of Figure \ref{Algo}.V\_5. 

\begin{algorithm}[htpb]                      
\rule{11.7cm}{0.01cm}
\caption{Check\_V\_5 ($r$ , $G$) -- $O(m^4)$} 
\rule{11.7cm}{0.01cm}
\\
{\bf Input:} a row $r$, the row-column intersection graph $G$.\\
{\bf Assumption:}  $r$ is not contained in a MCS of size $3$ or $4$.\\
{\bf Output:} returns a MCS ${\cal S}$ of size $5$ given by a forbidden 
induced subgraph of the form V containing $r$ if such a MCS exists, otherwise returns  "NO''.
\rule{11.7cm}{0.01cm}
\begin{algorithmic}[1]                
\FOR{any quadruplet $(r_i,r_j,r_k,r_l)$ of black vertices such that 
 $r,r_i,r_j,r_k,r_l$ are pairwise connected, except for one edge 
 $(r_a,r_b)$  in $\{r,r_i,r_j,r_k,r_l\}\times\{r,r_i,r_j,r_k,r_l\}$ missing}
\IF{$\{r_i,r_j,r_k,r_l\}$ is C1P}
\FOR{any pair $(a,b)$ in $(\{r,r_i,r_j,r_k,r_l\}-\{r_a,r_b\})
\times(\{r,r_i,r_j,r_k,r_l\}-\{r_a,r_b\})$}

\IF{$(a \cap b) - \cup \big(\{r,r_i,r_j,r_k,r_l\}-\{a,b\}\big) \neq \emptyset$, and $(r_k \cap a) - \cup \big(\{r,r_i,r_j,r_k,r_l\}-\{r_k,a\}\big) \neq \emptyset$, and $(r_l \cap b) - \cup \big(\{r,r_i,r_j,r_k,r_l\}-\{r_l,b\}\big) \neq \emptyset$}
\STATE  return $\{r,r_i,r_j,r_k,r_l\}$
\ENDIF
\ENDFOR
\ENDIF
\ENDFOR
\STATE  return "NO''
\end{algorithmic}
\rule{11.7cm}{0.01cm}
\end{algorithm}

\begin{proposition}
Algorithm $\mbox{Check\_V\_5}$ is correct and runs in $O(m^4)$ time.
\end{proposition}
\begin{proof}
Algorithm \mbox{Check\_V\_5} looks for an induced subgraph with 
$5$ black vertices $\{r,r_i,r_j,r_k,r_l\}$, that are  pairwise connected, 
except for one missing edge $(r_a,r_b)$  
in $\{r,r_i,r_j,r_k,r_l\}\times\{r,r_i,r_j,r_k,r_l\}$.
The $4$ black vertices that belong to the set with $r$, should correspond to
a set of rows that is C1P. Moreover, there should exist two particular 
rows (black vertices) of the set, with three columns (white vertices)
that satisfy the conditions on line 4 of the algorithm in order to fit 
the configuration depicted in Figure \ref{Algo}.V\_5.

Next, all the tests performed by 
Algorithm \mbox{Check\_V\_5} (lines 2-8 of the algoritm) on a given quatruplet 
$(r_i,r_j,r_k,r_l)$ are achieved in $O(1)$ thanks to the precomputations, 
and given $r$ there might be $O(m^4)$ such triplets.
Thus, Algorithm \mbox{Check\_V\_5} runs in $O(m^4)$ time.
\end{proof}

\subsubsection{MCS of size larger than $5$}

A MCS of the form V of size larger than $5$ contains exactly two kernels.
Depending on whether $r$ is a kernel or not, we distinguish two cases.\\

{\bf Case 1: If row $r$ is a kernel of the MCS}

Algorithm \mbox{Check\_V$_k$} recovers a MCS ${\cal S}$ of the form V of size 
larger than $5$ containing $r$ as a kernel, with the assumption that  $r$ 
is not contained in a MCS of size $3$, or $4$ (Figure \ref{Algo}.V$_k$).
The principle of the algorithm is similar to Algorithm \mbox{Check\_IV$_k$}. 
It relies in first choosing the second kernel $a$ of the MCS, and the column 
$c$, of the induced subgraphof type V responsible for ${\cal S}$, that is 
contained in both $r$ and $a$, but in no other row of the MCS  
(see Figure \ref{Algo}.V$_k$).
Next, it considers the subgraph $H$ of $G$ induced by the set of black 
vertices (rows) that are neighbors of $r$ and $a$, but do not contain 
$c$. We denote this subgraph by $H = G[N(r,a)-L(c)]$. Then, it looks for 
a set of rows $Q$, constituting a chordless path in $H$,
 such that $\{r\}\cup Q$ is a MCS of the form V.

\begin{algorithm}[htpb]                      
\rule{11.7cm}{0.01cm}
\caption{Check\_V$_k$ ($r$, $G$) -- $O(n^2m^2)$} 
\rule{11.7cm}{0.01cm}
\\
{\bf Input:} a row $r$, the row-column intersection graph $G$.\\
{\bf Assumption:}  $r$ is not contained in a MCS of size $3$, or $4$.\\
{\bf Output:} returns a MCS ${\cal S}$ of size larger that $5$ given by a 
forbidden induced subgraph of the form V such that $r$ is one of its kernel,
if such a MCS exists, otherwise returns  "NO''.
\rule{11.7cm}{0.01cm}
\begin{algorithmic}[1]
\FOR{ any black vertex $a\in N(r)$}
\FOR{ any column $c \in (r\cap a)$}
\STATE $H = G[N(r,a)-L(c)]$
\FOR{any connected component $C$ of $H$}
\STATE pick a a couple $(r_i,r_j)$ of black vertices in $C$ that satisfies 
1) $r_i$ and $r_j$ are not connected, and 2) $(r_i\cap r) - a \neq \emptyset$, 
and 3) $(r_j\cap a) - r \neq \emptyset$.
\STATE find a chordless path $P$ in $C$ linking $r_i$ and $r_j$
\STATE pick the smallest subpath $Q$ of $P$ linking two vertices $r'_i$ and 
$r'_j$, such that the couple $(r'_i,r'_j)$ also satisfies 1) and 2) and 3)
\STATE return $\{r\}\cup Q$
\ENDFOR
\ENDFOR
\ENDFOR
\STATE return ``NO''
\end{algorithmic}
\rule{11.7cm}{0.01cm}
\end{algorithm}

\begin{proposition}
Algorithm $\mbox{Check\_V}_c$ is correct and runs in $O(n^2m^2)$ time.
\end{proposition}
\begin{proof}
The proofs are similar to the proofs for the correctness and the
complexity of  Algorithm $\mbox{Check\_IV}_c$ as the two algorithms
are based on the same principle. However, here the complexity is 
multiplied by a factor $n$ due to considering all black vertices 
$a\in N(r)$.
\end{proof}

{\bf Case 2: If row $r$ is not a kernel of the MCS}

Algorithm Check\_V$_p$ recovers a MCS {\cal S} of the form V of size 
larger than $5$ containing r, but not as a kernel, with the assumptions 
that $r$ is not contained in a MCS of size $3$ or $4$, and r does not 
belong to an induced chordless cycle of $G_R$ (Figure \ref{Algo}.V). 

The principle of the algorithm is similar to the principle 
of Algorithm Check\_IV$_p$.
It consists in first choosing the two kernels $(a,b)$ of {\cal S} among the 
black vertices (rows) neighbors of $r$, and the column $c$, of the 
induced subgraph responsible for {\cal S}, that is contained in both
$a$ and $b$, but in no other row of the MCS. Next, the algorithm calls 
Algorithm Check\_V to look for the MCS  {\cal S} with $r$, $(a,b)$, $c$, 
and $G$ given as parameters.

\begin{algorithm}[htpb]                      
\rule{11.7cm}{0.01cm}
\caption{Check\_V$_p$ ($r$, $G$) -- $O(nm^7)$} 
\rule{11.7cm}{0.01cm}
\\
{\bf Input:} a row $r$, the row-column intersection graph $G$.\\
{\bf Assumption:}  $r$ is not contained in a MCS of size $3$, $4$.\\
$~~~~~~~~~~~~~~~~~~~~~r$ does not belong to an induced chordless cycle of $G_R$.\\
{\bf Output:} returns a MCS ${\cal S}$ of size larger that $5$ given by a 
forbidden induced subgraph of the form V containing $r$, but not as a kernel,
if such a MCS exists, otherwise returns  "NO''.
\rule{11.7cm}{0.01cm}
\begin{algorithmic}[1]                
\FOR{ any couple of connected black vertices $(a,b)\in N(r)^2$}
\FOR{ any column $c\in (a\cap b)-r$}
\STATE return Check\_V ($r$, $(a,b)$, $c$, $G$)
\ENDFOR
\ENDFOR
\STATE return ``NO''
\end{algorithmic}
\rule{11.7cm}{0.01cm}
\end{algorithm}

Algorithm Check\_V is called in Algorithm Check\_V$_p$. It recovers a MCS
${\cal S}$ of the form V of size larger than 5 containing $r$, given the 
row $r$,  the kernels $a$ and $b$ of the MCS, and the column $c$, of the 
induced subgraph responsible for ${\cal S}$, that is contained in $a$ and 
$b$, but in no other row of the MCS.

\begin{algorithm} [htpb]                   
\rule{11.7cm}{0.01cm}
\caption{Check\_V ($r$, $(a,b)$, $c$, $G$)-- $O(m^5)$}
\rule{11.7cm}{0.01cm}
\\
{\bf Input:} three rows $r$, $a$ and $b$, and a column $c\in a\cap b$ such that
 $r \in (N(a,b)-L(c))$ .\\
{\bf Assumption:}  $r$ is not contained in a MCS of size $3$, $4$ or $5$.\\
$~~~~~~~~~~~~~~~~~~~~~r$ is not contained in a MCS of type $IV$.\\
$~~~~~~~~~~~~~~~~~~~~~r$ does not belong to an induced chordless cycle of $G_R$.\\
{\bf Output:} returns a MCS ${\cal S}$ of size larger that $5$ given by a 
forbidden induced subgraph of the form V containing $a$, $b$, and $r$, and 
whose kernels are $a$ and $b$, if such a MCS exists, otherwise returns  "NO''.
\rule{11.7cm}{0.01cm}
\begin{algorithmic}[1] 
\STATE $H = G[N(a,b)-L(c)]$
\STATE let $C=(V_C,E_C)$ be the connected component of $H$ to which $r$ belongs.
\STATE let $V_A$ be the set of vertices $V_A= \{u\in V_C ~:~ (u\cap a)-b \neq \emptyset \}$.
\STATE let $V_B$ be the set of vertices $V_B= \{v\in V_C ~:~ (v\cap b)-a \neq \emptyset \}$.
\STATE let $E_{AB}$ be the set of edges $E_{AB}= \{(u,v), u \in V_A, v \in V_B ~:~  u\cap v = \emptyset\}$.
\STATE let $V_a$ be the set of vertices  $V_a= \{u\in V_C ~:~ u-a \neq \emptyset\}$, and $E_a$ be the set of edges $E_a= \{(u,v)\in V_a^2 ~:~  u\cap v = \emptyset\}$.
\STATE let $V_b$ be the set of vertices  $V_b= \{u\in V_C ~:~ u-b \neq \emptyset\}$, and $E_b$ be the set of edges $E_b= \{(u,v)\in V_b^2 ~:~  u\cap v = \emptyset\}$.
\STATE let $D=(V_D,E_D)$ be the graph such that $V_D = V_C$ and $E_D=E_C\cup E_{AB}\cup E_a\cup E_b$ 
\STATE $Q$ =  Check\_I ($r$, $D_R$)
\IF{ $Q \neq$ "NO''}
\STATE return $\{a,b\}\cup Q$
\ENDIF
\STATE return "NO''
\end{algorithmic}
\rule{11.7cm}{0.01cm}
\end{algorithm}

\begin{proposition}
Algorithm $\mbox{Check\_V}_p$ is correct and runs in $O(nm^7)$ time.
\end{proposition}
\begin{proof}
In order to prove the correctness and the complexity of Algorithm 
$\mbox{Check\_V}_p$, we need to prove the correctness and give the
complexity of Algorithm $\mbox{Check\_V}$ that is called in 
$\mbox{Check\_V}_p$.

The correctness of Check\_V comes from the fact that $r$ does not
belong to any chordless cycle in the graph $C$ computed at line 2 of the
algorithm by assumption. Let $Q$ be a chordless cycle in the graph $D$ 
containing vertex $r$, computed at line 9 of the algorithm. 
Since $r$ does not belong to an induced chordless cycle of the $C$ by 
assumption, then $Q$ necessarily contains at least one edge belonging to 
the set $E_{AB}\cup E_a\cup E_b$.

\noindent
We first give two trivial but useful properties for the remaining of the proof:
\begin{itemize}
\item[(i)] For any two edges of $Q$, there always exists two extremities $u$
 and $v$ of these edges, one in each edge, that are not disjoint in the graph 
 $C$, 
i.e $u\cap v = \emptyset$ 
\item[(ii)] $V_A \subseteq V_b$, and  $V_B \subseteq V_a$.
\end{itemize}

\noindent
We also prove the following useful property: 
\begin{itemize}
\item[(iii)] $V_a \subseteq (V_B\cup V_b)$ and $V_b \subseteq (V_A\cup
  V_a)$. Let $x \in V_a,$ there exists $c$ such that $c \in x$ and
   $c \not \in a.$ Then, either $c \not \in b$ in
  which case $x \in V_b$, or $c \in b,$ which implies
  that $x \in V_B$. The proof is similar for  $V_b \subseteq (V_A\cup
  V_a).$
\end{itemize}

We now prove that the cycle $Q$ necessarily contains at most one edge of
the set $E_{AB}\cup E_a\cup E_b$. Indeed, if $Q$ contains two edges of
$E_{AB}\cup E_a\cup E_b$, let $u,v$ be two disjoint extremities of these edges
(Property (i)). We can distinguish $7$ cases according to the belonging
of $u$ and $v$ to the sets $V_A$, $V_B$, $V_a$ and $V_b$, and we show 
in the following that, in all these cases, a chord is induced in the
chordless cycle $Q$ in the graph $D$: contradiction.

\begin{enumerate}
\item If $(u,v) \in V_A^2$ (resp. $(u,v) \in V_B^2$), then from Property (ii), 
$(u,v) \in V_b^2$ (resp.  $(u,v) \in V_a^2$), and thus $(u,v) \in E_b$
(resp. $(u,v) \in E_a$).
\item If $(u,v) \in V_a^2$ (resp.  $(u,v) \in V_b^2$), then  $(u,v) \in E_a$ (resp. $(u,v) \in E_b$).
\item If $(u,v) \in V_A\times V_B$ (or the symmetric), then  $(u,v) \in E_{AB}$.
\item If $(u,v) \in V_A\times V_a$ (or the symmetric), then  from Property (iii),
$(u,v) \in V_A\times V_B$ or $(u,v) \in V_A\times V_b$, and thus $(u,v) \in E_{AB}$ or $(u,v) \in E_b$ from cases 3. and 6.
\item If $(u,v) \in V_B\times V_b$ (or the symmetric), then  from Property (iii), 
$(u,v) \in V_B\times V_A$ or $(u,v) \in V_B\times V_a$, and thus $(u,v) \in E_{AB}$ or $(u,v) \in E_a$ from cases 3. and 6.
\item If $(u,v) \in V_A\times V_b$ (resp.  $(u,v) \in V_B\times V_a$) (or the symmetric), then from Property (ii),  $(u,v) \in V_b^2$ (resp.  $(u,v) \in V_a^2$), and thus $(u,v) \in E_b$ (resp. $(u,v) \in E_a$).
\item If $(u,v) \in V_a\times V_b$ (or the symmetric), then from Property (iii), 
$(u,v) \in V_b\times V_b$ or $(u,v) \in V_B\times V_b$, and thus $(u,v) \in E_b$ or $(u,v) \in E_{AB}\cup E_a$ from cases 1 and 5.
\end{enumerate}

\noindent
In consequence, there exits at most one edge, and then exactly one edge 
of the set $E_{AB}\cup E_a\cup E_b$ in the cycle $Q$ in the graph $D$.
Next, let $(r_i,r_j)$ be the only edge of $Q$ belonging to 
$E_{AB}\cup E_a\cup E_b$.
We show that  $(r_i,r_j)\not\in E_a\cup E_b$. Indeed, if $(r_i,r_j)\in E_a$
(resp. $(r_i,r_j)\in E_b$), then the set $\{a\}\cup Q$ (resp. $\{b\}\cup Q$) 
satisfies the conditions
to be a MCS of type IV with $a$ (resp. $b$) as kernel, which is impossible 
by assumption.

So, we have $(r_i,r_j)\in E_{AB} - (E_a\cup E_b)$. Finally, removing the edge  
$(r_i,r_j)$ from the cycle yields a chordless path $Q$ in $G$ 
containing $r$ such that each vertex of $Q$ 
is connected to vertices  $a$ and $b$, and the extremities $r_i$ and $r_j$ of 
$Q$ satisfy 1) $r_i$ and $r_j$ are not connected, and 2) 
$(r_i\cap a) - b \neq \emptyset$, and 3) $(r_j\cap b) - a \neq \emptyset$.
and 4) $Q$ does not contain any smaller subpath satisfying conditions 
1) and 2) and 3). 
These conditions are necessary and sufficient for the 
set $\{a,b\} \cup Q$ to form the rows of an induced subgraph of the form V, 
and this set cannot contain a smaller MCS since such a MCS would be:
\begin{itemize}
 \item either a MCS of size $3$ including $a$ or $b$, 
 \item or a MCS of type II or III necessarily including $a$ or $b$ as kernel,
 \item or a MCS of  type IV and size larger than $3$  having $a$ or $b$ as 
kernel.
 \item or a MCS of  type IV and size larger than $3$  having $a$ and $b$ as 
kernels.
\end{itemize}
The 3 cases are impossible,  since they would induce a chord from the
set $E_{AB}\cup E_a\cup E_b$ in the chordless cycle induced by $Q$ in 
the graph $D$.

The correctness of Algorithm Check\_V$_p$ follows
immediately from the correctness of Algorithm Check\_V.

Algorithm $\mbox{Check\_V}$ calls Algorithm $\mbox{Check\_I}$. 
Both algorithms have the same time complexity in $O(m^5)$ time.
It follows immediately that  Algorithm $\mbox{Check\_IV}_p$
runs  in $Onm^7$ time.
\end{proof}

\paragraph{Acknowledgment} We would like to thanks Nicolas Trotignon for 
his valuable comments on induced subgraphs and also Juraj Stacho for
his participation to some meeting on the subject.

\bibliographystyle{plain}
\bibliography{consecutives}
\end{document}